\documentclass[prc,twocolumn,groupedaddress]{revtex4}
\usepackage{graphicx,amsfonts,color,epsfig}

\newcommand{\SU}[1]{\ensuremath{\mathrm{SU}( #1 )}}
\newcommand{\Un}[1]{\ensuremath{\mathrm{U}( #1 )}}
\newcommand{\SO}[1]{\ensuremath{\mathrm{SO}( #1 )}}

\newcommand{\SpR}[1]{\ensuremath{\mathrm{Sp}( #1,\mathbb{R} )}}
\newcommand{\su}[1]{\ensuremath{\mathfrak{su}( #1 )}}

\newcommand{\so}[1]{\ensuremath{\mathfrak{so}( #1 )}}

\newcommand{\spR}[1]{\ensuremath{\mathfrak{sp}( #1, \mathbb{R} )}}
\newcommand{\ket}[1]{\ensuremath{\left| #1 \right\rangle}}

\newcommand{\betb}{\begin{tabular}{p{4.0cm}p{9.0cm}}}
\newcommand{\entb}{\end{tabular}}

\newcommand{\ho}{\ensuremath{\hbar\Omega}}
\newcommand{\ph}[1]{\ensuremath{#1}p-\ensuremath{#1}h}
\begin{document}

\centerline{ \hskip 9.3in
\begin{tabular}{r}
UCRL-JRNL-227466 \\
SLAC-PUB-12386
\end{tabular}
}

\title{Symplectic Symmetry and the {\it Ab Initio} No-Core Shell Model}
\author{Jerry P. Draayer}
\author{Tom\'a\v{s} Dytrych}
\author{Kristina D. Sviratcheva}
\author{Chairul Bahri}
\affiliation{Department of Physics and Astronomy, Louisiana State
University, Baton
Rouge, LA 70803, USA}
\author{James P. Vary}
\affiliation{Department of Physics and Astronomy, Iowa State
University, Ames, IA 50011,
USA}
\affiliation{Lawrence Livermore National Laboratory, L-414, 7000 East
Avenue, Livermore,
California, 94551, USA}
\affiliation{Stanford Linear Accelerator Center, MS81, 2575 Sand Hill
Road, Menlo Park,
California, 94025, USA}

\begin{abstract}
The symplectic symmetry of eigenstates for the $0^+_{gs}$ in $^{16}$O and the $0^+_{gs}$ and lowest 
$2^+$ and $4^+$ configurations of $^{12}$C that are
well-converged within the framework of the no-core shell model with the 
JISP16 realistic interaction is examined. These states
are found to project at the $85-90$\% level onto very few symplectic representations
including the most deformed configuration,
which confirms the importance of a symplectic no-core shell model and 
reaffirms the relevance of the Elliott \SU{3}
model upon which the symplectic scheme is built.
\end{abstract}

\maketitle

\section{Introduction}

With the development of high-precision nucleon-nucleon ({\it NN}) interactions derived from meson
exchange theory and the latest advances 
in chiral effective field theory based on QCD that provide
components of nuclear forces that can be matched to
the underlying theory of quarks and gluons, {\it ab-initio} nuclear theoretical models play a crucial
role towards a deeper understanding of fundamental aspects of nuclear physics. 
{\it Ab-initio} calculations target reproducing nuclear structure features while employing
realistic {\it NN} (or many-nucleon) interactions and hence bridge from nuclear structure
considerations to the nucleon constituent degrees of freedom and, in turn, to astrophysical
phenomena, including nucleosynthesis and neutron stars, as well as towards a further exploration of
$3N$ nuclear forces and exotic physics of rare isotopes.   

The {\it ab-initio} No-Core Shell Model (NCSM) \cite{NCSM} with modern realistic
interactions yields a  good description of the low-lying states in few-nucleon systems
\cite{NavratilBK98_00} as
well as in more complex nuclei like $^{12}$C \cite{NCSM}. In
addition to advancing our understanding of the propagation of the nucleon-nucleon
force in nuclear matter and clustering phenomena
\cite{KarataglidisDAS95,FunakiTHSR03}, modeling the structure of $^{12}$C, $^{16}$O and
similar nuclei is also important for gaining a better understanding
of other physical processes such as parity-violating electron scattering from light nuclei
\cite{MusolfD92}
and results gained through neutrino studies 
\cite{HayesNV03} as well as for making
better predictions for
capture reaction rates that figure prominently, for example, in the
burning of He in
massive stars \cite{Brune99}.

Our investigations  show that the $0^{+}_{gs}$ and the
lowest $2^{+}$ and $4^{+}$ states in $^{12}$C and the $^{16}$O ground state reflect the presence of
an underlying
symplectic \spR{3} algebraic structure \footnote{We use lowercase (capital) letters for algebras
(groups).}.  This is achieved through the projection of realistic NCSM eigenstates onto 
\SpR{3}-symmetric basis states of the symplectic shell model that are free of
spurious center-of-mass motion.
Typically, eigenstates of the NCSM are reasonably well converged in the
$N_{max}=6$ (or
6\ho) basis space with an effective interaction based on the JISP16 realistic
interaction \cite{ShirokovMZVW04}. In particular, calculated binding
energies as well as  other observables for $^{12}$C
such as B(E2;$2^{+}_{1}\!\rightarrow\!0^{+}_{gs}$),
B(M1;$1^{+}_{1}\!\rightarrow\!0^{+}_{gs}$), ground-state proton rms
radii and the
$2^{+}_{1}$ quadrupole moment all lie reasonably close to the
measured values. 
The symplectic shell model itself \cite{Sp3R1,Sp3R2} is a microscopic realization
of the successful Bohr-Mottelson collective model. It is also a multiple oscillator
shell generalization of Elliott's \SU{3} model. Hence, this analysis
provides, for the
first time, a close examination of symmetries in nuclei as unveiled
through {\it ab-initio} calculations of the NCSM type with realistic interactions.

The rotational nature of the ground-state band in $^{12}$C nucleus has long been
recognized, and early-on group-theoretical methods with \SU{3} serving as the
underpinning (sub-)symmetry were used in its description. 
Symplectic algebraic
approaches have achieved a very good reproduction of low-lying energies and B(E2) values in light
nuclei \cite{RosensteelR80,DraayerWR84} and specifically in $^{12}$C
 using phenomenological interactions \cite{EscherL02} or
truncated symplectic
basis with simplistic (semi-) microscopic interactions
\cite{ArickxBD82,AvanciniP93}.
Here, we establish the dominance of the symplectic \SpR{3} symmetry
in the {\it ab-initio} NCSM wavefunctions.
This in turn opens up a new and exciting possibility
for representing significant high-$\hbar \Omega $ collective modes 
by extending the NCSM basis space beyond its current limits 
through \SpR{3} basis states, 
which yields a dramatically smaller basis space to achieve convergence
of higher-lying collective modes.


\section{Symplectic Shell Model}
The symplectic shell model is based on the noncompact symplectic
\spR{3} algebra
that with its subalgebraic structure unveils the underlying physics
of a microscopic
description of collective modes in nuclei \cite{Sp3R1,Sp3R2}. The latter follows from
the fact that the
mass quadrupole and monopole moment operators, the many-particle
kinetic energy, the
angular and vibrational momenta are all elements of the $\spR{3}\supset \su{3}
\supset\so{3}$  algebraic structure. Hence, collective states of a nucleus with
well-developed quadrupole and monopole vibrations as well as
collective rotations are
described naturally in terms of irreducible representations (irreps)
of \SpR{3}.
Furthermore, the elements of the \spR{3} algebra are constructed as
bilinear products
in the harmonic oscillator (HO) raising and lowering 
operators that in turn are expressed
through particle
coordinates and linear momenta. This means the basis states of a
\SpR{3} irrep can be expanded in a HO ($m$-scheme) basis,
the same basis used in the NCSM, thereby facilitating symmetry identification. 

The symplectic basis states are labeled (in standard notation \cite{Sp3R1,Sp3R2})
according to the reduction chain
\begin{equation}
\begin{tabular}{ccccc}
$\SpR{3}$ & ~$\supset $ & $~~~\Un{3}$  & $\supset$  &    ~~~$\SO{3}$ \\
~~$\Gamma_\sigma\;\;$& $\Gamma_n\rho$ & ~~~$\Gamma_\omega$ &  $\kappa$  & ~~$L$
\end{tabular}
\end{equation}
and are constructed by acting with polynomials $\mathcal{P}$ in the
symplectic raising
operator, $A^{(2\,0)}$, on a set of basis states of the symplectic
bandhead, $\ket{\Gamma_{\sigma}}$, which is a \SpR{3} lowest-weight
state
($B^{(0\,2)}|\Gamma_\sigma\rangle = 0$, where the symplectic
lowering operator $B^{(0\,2)}$ is the adjoint of $A^{(2\,0)}$); that
is,
\begin{equation}
|\Gamma_\sigma \Gamma_n\rho\Gamma_\omega \kappa (LS) J M_{J}\rangle 
{\textstyle 
= \left[{\mathcal{P}^{\Gamma_n}(A^{(2\,0)})
\times 
|\Gamma_\sigma}\rangle\right]^{\rho\Gamma_\omega}_{\kappa 
(LS) J M_{J}},
}
\label{bs}
\end{equation}
where $\Gamma_\sigma$ $\equiv $ $N_\sigma\left(\lambda_{\sigma}\, \mu_{\sigma}\right)$
labels \SpR{3} irreps with
$\left(\lambda_{\sigma}\,\mu_{\sigma}\right)$ denoting a \SU{3}
lowest-weight state,
$\Gamma_{n}\equiv n\left(\lambda_{n}\, \mu_{n}\right)$, and
$\Gamma_\omega\equiv N_\omega\left(\lambda_{\omega}\,
\mu_{\omega}\right)$.
The $\left(\lambda_{n}\, \mu_{n}\right)$ set
gives the
overall \SU{3} symmetry of
$\frac{n}{2}$ coupled raising 
operators in $\mathcal{P}$, $\left(\lambda_{\omega}\,
\mu_{\omega}\right)$ specifies the \SU{3} symmetry of the
symplectic state, and $N_\omega=N_\sigma+n$ is the total number of
oscillator quanta related to the eigenvalue,
$N_\omega \hbar\Omega$, of a HO Hamiltonian that is free of 
spurious modes. Consequently, the symplectic basis states bring forward important information about
the nuclear shape deformation  in terms of the
\SU{3} labels,
$(\lambda_\omega\ \mu_\omega)$, for example,
$(0\ 0)$,
$(\lambda\ 0)$ and $(0\ \mu)$  describe spherical, prolate and oblate shapes, respectively.

The symplectic raising operator $A^{(2\,0)}_{lm}$, 
which is a \SU{3} tensor 
with $\left(\lambda\, \mu\right)=\left(2\,0\right)$  character,
can be expressed as a bilinear product of the HO raising operators,
\begin{equation}
{\textstyle
A^{(20)}_{lm}= \frac{1}{\sqrt{2}} \sum_{i}
\left[b_{i}^{\dagger}\times b_{i}^{\dagger}\right]^{(20)}_{lm} 
- \frac{1}{\sqrt{2}A} \sum_{s,t}
\left[b^{\dagger}_{s}\times b^{\dagger}_{t}\right]^{(20)}_{lm},
}
\label{A20}
\end{equation}
where the sums are over all $A$ particles of the system. The first term in (\ref{A20})
describes 2\ho~ one-particle-one-hole (1p-1h) excitations (one particle raised by two shells) and the
second term eliminates the spurious center-of-mass excitations in the construction (\ref{bs}). 
For the purpose of comparison to NCSM results, the basis states of the 
$\ket{\Gamma_{\sigma}}$ bandhead
in (\ref{bs}) are constructed in a $m$-scheme basis,
\begin{eqnarray}
&&\ket{\Gamma_{\sigma}\kappa (L_0S_0) J_0 M_{0}} {\textstyle =} \nonumber \\
&&\left[\mathcal{P}^{(\lambda_{\pi}\,\mu_{\pi})}_{S_{\pi}}(a^{\dagger}_{\pi})
{\textstyle \times}
\mathcal{P}^{(\lambda_{\nu}\,\mu_{\nu})}_{S_{\nu}}(a^{\dagger}_{\nu})
\right]^{(\lambda_{\sigma}\,\mu_{\sigma})}_{\kappa (L_0S_0) J_0 M_{0}}\ket{0},
\label{bandhead}
\end{eqnarray}
where \ket{0} is a vacuum state,
$\mathcal{P}^{(\lambda_{\pi}\,\mu_{\pi})}_{S_{\pi}}$ and
$\mathcal{P}^{(\lambda_{\nu}\,\mu_{\nu})}_{S_{\nu}}$ denote polynomials of
proton ($a^{\dagger}_{\pi}$) and neutron ($a^{\dagger}_{\nu}$) creation
operators coupled to good
\SU{3}$\times$\SU{2} symmetry.


\section{Results and Discussions}
The lowest-lying eigenstates of $^{12}$C and $^{16}$O were calculated using the
NCSM as implemented through the
Many Fermion Dynamics ~(MFD) code \cite{Vary92_MFD} with an
effective interaction derived from the realistic JISP16 {\it NN}
potential \cite{ShirokovMZVW04} for different \ho~ oscillator strengths. 
We are particularly interested in the $^{16}$O ground state and the $J\!=\!0^{+}_{gs}$ and the
lowest $J\!=\!2^{+}(\equiv\!2^{+}_{1})$ and $J\!=\!4^{+}(\equiv\!4^{+}_{1})$ states of the
ground-state (gs) rotational band in $^{12}$C.
\begin{table}[th]
\caption{Probability distribution of NCSM eigenstates for $^{12}$C across the 
dominant \ph{0} and 2\ho~\ph{2} \SpR{3} irreps, \ho=15 MeV.\label{TABLE_15MeV}}
\begin{ruledtabular}
\begin{tabular}{lcrrrrr}
& & $0\hbar\Omega$ & $2\hbar\Omega$ & $4\hbar\Omega$ & $6\hbar\Omega$ & Total \\
\hline
\multicolumn{7}{c}{$J=0$} \\
\hline
$\SpR{3}$          & $(0\;4)S=0$     & $46.26$ & $12.58$ & $4.76$  & $1.24$  & $64.84$\\
                   & $(1\;2)S=1$     & $4.80$  & $2.02$  & $0.92$  & $0.38$  & $8.12$\\
   	               & $(1\;2)S=1$ & $4.72$  & $1.99$  & $0.91$  & $0.37$  & $7.99$ \\
            & 2\ho~\ph{2}            &         & $3.46$  & $1.02$  & $0.35$  & $4.83$\\
\cline{2-7}
                  & Total       & $55.78$& $20.05$& $7.61$ &  $2.34$  & $85.78$ \\
NCSM              &            & $56.18$ & $22.40$ & $12.81$ &  $7.00$  & $98.38$ \\
\hline
\multicolumn{7}{c}{$J=2$} \\
\hline
$\SpR{3}$          & $(0\;4)S=0$ & $46.80$ & $12.41$ & $4.55$  & $1.19$     &  $64.95$\\
                   & $(1\;2)S=1$ & $4.84$  & $1.77$  & $0.78$  &  $0.30$    &  $7.69$ \\
                   & $(1\;2)S=1$ & $4.69$  & $1.72$  & $0.76$  &  $0.30$    &  $7.47$ \\
                   & 2\ho~\ph{2} &         & $3.28$  & $1.04$  &  $0.38$    &  $4.70$\\
\cline{2-7}
     &    Total     & $56.33$ & $19.18$ & $7.13$ &  $2.17$  & $84.81$ \\
NCSM    &           & $56.18$ & $21.79$ & $12.73$ &  $7.28$ & $98.43$ \\
\hline
\multicolumn{7}{c}{$J=4$} \\
\hline
$\SpR{3}$          & $(0\;4)S=0$ & $51.45$ & $12.11$ & $4.18$   & $1.04$     & $68.78$\\
                   & $(1\;2)S=1$ & $3.04$  & $0.95$  & $0.40$   & $0.15$     & $4.54$ \\
                   & $(1\;2)S=1$ & $3.01$  & $0.94$   & $0.39$  & $0.15$     & $4.49$ \\
                  & 2\ho~\ph{2}  &         & $3.23$   & $1.16$  & $0.39$     & $4.78$\\
\cline{2-7}
       &    Total & $57.50$ & $17.23$ & $6.13$  &  $1.73$  & $82.59$ \\
NCSM   &          & $57.64$ & $20.34$ & $12.59$ &  $7.66$  & $98.23$
\end{tabular}
\end{ruledtabular}
\end{table}
\begin{table}[th]
\caption{Probability distribution of the NCSM eigenstate for the $J=0$ ground state in
$^{16}$O across the \ph{0} and dominant 2\ho~\ph{2} \SpR{3} irreps, \ho=15 MeV.
\label{TABLE_O16}}
\begin{ruledtabular}
\begin{tabular}{lcrrrrr}
& & $0\hbar\Omega$ & $2\hbar\Omega$ & $4\hbar\Omega$ & $6\hbar\Omega$ & Total \\
\hline
$\SpR{3}$ & $(0\;0)S=0$ & $50.53$ & $15.87$ & $6.32$ & $2.30$ & $75.02$ \\
          & 2\ho~\ph{2} & & $5.99$ & $2.52$ & $1.32$ & $9.83$ \\
\cline{2-7}
                 & Total       & $50.53$ & $21.86$  & $8.84$  &  $3.62$ & $84.85$ \\
NCSM &                         & $50.53$ & $22.58$ & $14.91$ &  $10.81$ & $98.83$ \\
\end{tabular}
\end{ruledtabular}
\end{table}

\begin{figure*}[th]
\includegraphics[width=0.6\textwidth]{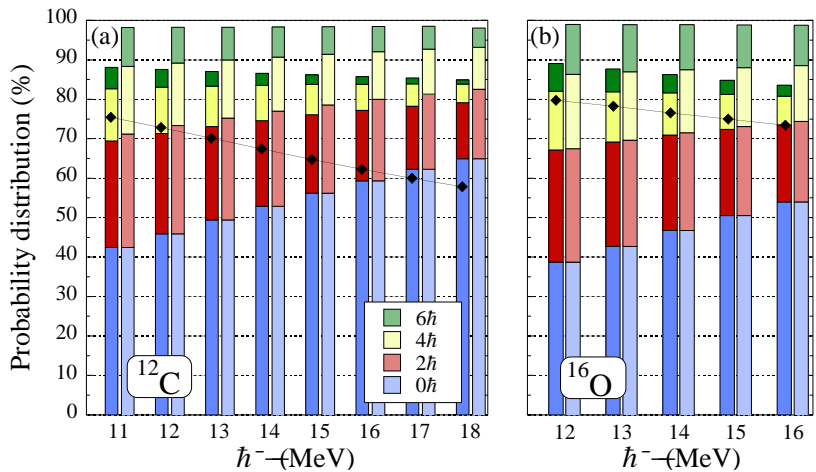}
\caption{
Ground $0^+$ state probability distribution 
over $0$\ho~ (blue, lowest) to $6$\ho~ (green, highest) subspaces
for the most dominant \ph{0} + 2\ho~\ph{2} \SpR{3} irrep case 
(left) and NCSM (right) together with the leading irrep
contribution (black diamonds), $(0~4)$ for $^{12}$C (a) and $(0~0)$ for $^{16}$O (b), as a function of
the
\ho~ oscillator strength, $N_{max}=6$.}
\label{C12prblty_vs_hw}
\end{figure*}
For both nuclei we constructed all of the \ph{0} and 2\ho~\ph{2} (2 particles raised by one 
shell each) symplectic bandheads and
generated their \SpR{3} irreps up to $N_{max}=6$ (6\ho~ model space).
The typical dimension of a symplectic irrep basis in
the $N_{max}=6$ space is on the order of $10^{2}$ as compared to
$10^{7}$ for the full NCSM $m$-scheme basis space.

Analysis of overlaps of the symplectic states with the NCSM eigenstates
for 2$\hbar\Omega$, 4$\hbar\Omega$, and 6$\hbar\Omega$ model spaces ($N_{max}=2,4,6$) reveals the
dominance of the
\ph{0} \SpR{3} irreps.  For the $0^+_{gs}$ and the lowest $2^+$ and
$4^+$ states in $^{12}$C  there are nonnegligible overlaps for only 3 of the 13 \ph{0} \SpR{3} irreps,
namely, the leading (most deformed) representation with $N_\sigma =24.5$ and
$(\lambda_\sigma~\mu_\sigma)=(0~4)$, and carrying spin
$S=0$ together with two $24.5~(1~2)$
$S=1$ irreps with different bandhead constructions for
protons and neutrons. For the ground state of $^{16}$O there is
only one possible \ph{0} \SpR{3} irrep, $34.5~(0\,0)$ $S=0$. In addition, among the 2\ho~\ph{2}
\SpR{3} irreps only a small fraction contributes significantly to the overlaps and it includes the  
most deformed symplectic bandhead configurations that correspond to oblate shapes in $^{12}$C and
prolate ones in $^{16}$O.

The overlaps of the most dominant symplectic states with the $^{12}$C and $^{16}$O NCSM eigenstates
under consideration in the $0$, $2$, $4$ and $6\hbar\Omega$ subspaces are 
given in Table~\ref{TABLE_15MeV} and \ref{TABLE_O16}.
In order to speed up the calculations, we retained only the largest amplitudes of the NCSM states,
those sufficient to account for at least 98\% of the norm which is quoted also in the table. 
The results show that approximately 85\% of the NCSM eigenstates for $^{12}$C
($^{16}$O) fall within a subspace spanned by the few most significant \ph{0} and 2\ho~\ph{2} \SpR{3}
irreps, with the 2\ho~\ph{2} \SpR{3} irreps accounting for 5\% (10\%) and with the leading irrep,
$(0~4)$ for $^{12}$C and $(0~0)$ for $^{16}$O, carrying close to 70\% (75\%) of the NCSM
wavefunction.

In addition, the projection of the NCSM wavefunctions onto the symplectic space slightly changes as
one varies the oscillator strength \ho~(see, e.g., Fig.
\ref{C12prblty_vs_hw} for the $0^+_{gs}$ state). The overall overlaps increase towards smaller \ho~HO
frequencies and, for example, for
$0^+_{gs}$ it is 90\% in the $N_{max}=6$ and $\ho=11$MeV case. Clearly, the largest contribution
comes from the leading, most deformed $(0~4)S=0$ \SpR{3} irrep for $^{12}$C and $(0~0)S=0$ for
$^{16}$O, growing to $\approx 90\%$ of the total
\SpR{3}-symmetric part for \ho~=11 MeV. As expected, Fig.
\ref{C12prblty_vs_hw} also confirms that with increasing $\hbar\Omega$ the higher \ho~ excitations
contribute less while the lower 0\ho~ configurations grow in importance. 

\begin{figure*}[t]
\centerline{\epsfxsize=4in\epsfbox{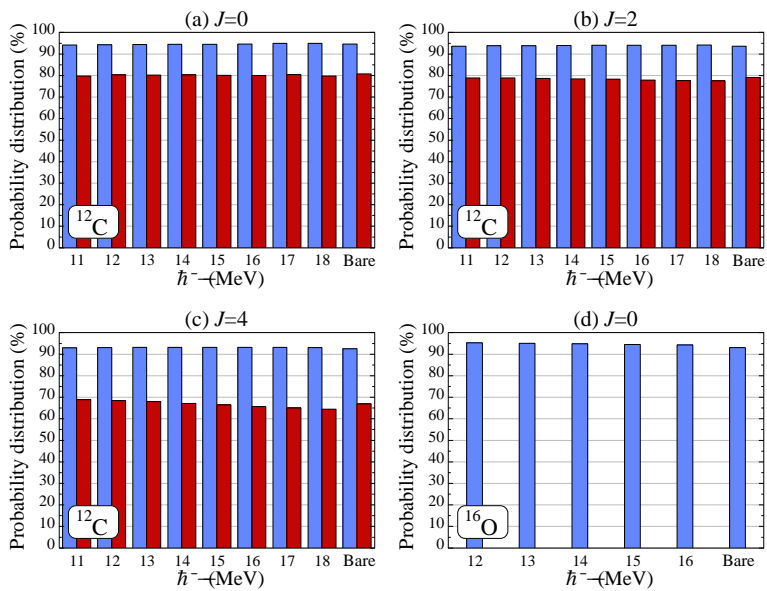}} 
\caption{
Projection of the $S=0$ (blue, left) [and $S=1$ (red, right)] \SpR{3} irreps onto the
corresponding significant spin components of the NSCM wavefunctions for (a) $0^{+}_{gs}$, (b)
$2^{+}_1$, and (c)
$4^{+}_1$ in
$^{12}$C and (d) $0^{+}_{gs}$ in $^{16}$O, for effective interaction for different
$\hbar\Omega$ oscillator strengths and bare interaction. (We present only the most significant spin
values that account for more than 90\% of a NCSM wavefunction).}
\label{Overlaps_SpinScaled}
\end{figure*}
Furthermore, the $S=0$ part of all three  NCSM eigenstates for $^{12}$C is almost entirely projected
(95\%) onto only six $S=0$ symplectic irreps included in Table
\ref{TABLE_15MeV}, with as much as 90\% of the spin-zero NCSM states accounted solely by 
the leading $(0\,4)$. 
The $S=1$ part is also remarkably well described by merely two \SpR{3} irreps. 
Similar results are observed for the ground state of $^{16}$O.
The outcome reveals an important property for the symplectic dynamical symmetry
reflected within the spin projections of the converged NCSM states, namely, their \SpR{3} symmetry
and hence the geometry of the nucleon system is independent of the \ho~oscillator strength
(Fig.~\ref{Overlaps_SpinScaled}). The symplectic  symmetry is equally present in the spin parts of
the NCSM wavefunctions for
$^{12}$C as well as $^{16}$O
regardless of whether the bare or the effective
interactions are used. This suggests that the  Lee-Suzuki transformation, which effectively
compensates for the finite space truncation by  renormalization of the bare interaction, does not
affect the \SpR{3} symmetry structure of the spatial wavefunctions. Hence,
the symplectic structure detected in the present analysis for $6\ho$ model space is what would emerge
in NSCM evaluations with a sufficiently large model space and bare interaction. 

\begin{figure}[th]
\includegraphics[width=0.4\textwidth]{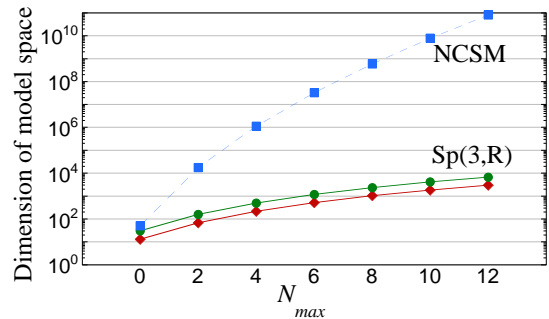}
\caption{Dimension of the NCSM (blue squares) and $J=0,2,$ and $4$ \SpR{3} 
(red diamonds for the 3 most significant \ph{0} irrep case and green
circles for when all 13 \ph{0} irreps are included) model spaces as a function of
maximum allowed \ho~ excitations, $N_{max}$, for $^{12}$C.}
\label{dimMdlSpace}
\end{figure}

As $N_{max}$ is increased the dimension of the $J=0,2,$ and
$4$ symplectic space  built on the \ph{0} \SpR{3} irreps
grows very slowly compared to the NCSM space dimension (Fig. \ref{dimMdlSpace}), which remains a
small fraction of the NCSM basis space even when the most dominant 2\ho~\ph{2} \SpR{3} irreps are
included [it is 1.29\% for the 2\ho~ model space,	$8.7\times 10^{-2}\%$ (4\ho),
$8.5\times 10^{-3}\%$ (6\ho),
$9.9\times 10^{-4}\%$ (8\ho),
$1.3\times 10^{-4}\%$ (10\ho), and
$1.6\times 10^{-5}\%$ (12\ho)]. The space reduction is even
more dramatic in the case of $^{16}$O for the most dominant \ph{0} and 2\ho~\ph{2}
configurations [0.88\% for the 2\ho~ model space,	$1.2\times 10^{-2}\%$ (4\ho),
$4.4\times 10^{-4}\%$ (6\ho),
$2.6\times 10^{-5}\%$ (8\ho),
$2.1\times 10^{-6}\%$ (10\ho), and
$2.1\times 10^{-7}\%$ (12\ho)]. This means that a space
spanned by a set of symplectic basis states may be computationally manageable even when high-\ho~
configurations are included. It is important to note that 2\ho~2p-2h (2 particles raised by one 
shell each) and higher rank \ph{n} excitations and allowed multiples thereof
can be included by building them into an expanded set of
lowest-weight
\SpR{3} starting state configurations. The same ``build-up'' logic, (\ref{bs}), holds
because by construction these additional starting state configurations are also
required
to be lowest-weight \SpR{3} states. 
Note that if one were to include all
possible lowest-weight \ph{n} starting state configurations $(n \leq
N_{max})$,
and allowed multiples thereof, one would span the entire NCSM space.

\begin{figure}[th]
\includegraphics[width=0.4\textwidth]{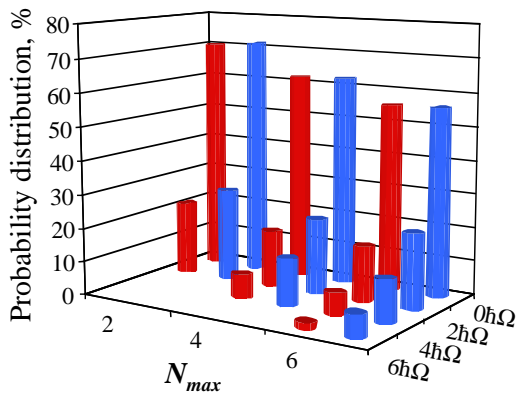}
\caption{
NCSM (blue, right) and \SpR{3} (red, left) probability distribution over $0$\ho~to
$N_{max}$ \ho~ subspaces for the $0^+_{gs}$ of $^{12}$C for
different model spaces, $N_{max}$, with $\hbar\Omega = 15$MeV.
}
\label{distribMdlSpace}
\end{figure}

Examination of the role of the model space truncation specified by $N_{max}\ho$
reveals that
the general features of all outcomes are retained as the space is expanded from
2$\hbar\Omega$ to 6$\hbar\Omega$ (see, e.g., Fig. \ref{distribMdlSpace} for
$0^+_{gs}$ in $^{12}$C). Specifically, the same three \SpR{3} irreps in $^{12}$C,
(0~4)$S=0$ and the two (1~2)$S=1$, dominate for all $N_{max}$ values
with the large overlaps of the NCSM eigenstates with the leading
symplectic irreps preserved, albeit distributed outward across
higher $\hbar\Omega$ 
excitations as the number of active shells increases.
In this regard, it may be interesting to understand the importance of
the latter beyond the $6\hbar \Omega $ model space and their role in
shaping other low-lying states in $^{12}$C and $^{16}$O such as the second
$0^+$. This task, albeit challenging, is feasible for the no-core
shell model with symplectic \SpR{3} extension and will be part
of a follow-on study.

The $0^{+}_{gs}$ and $2^{+}_1$ states in $^{12}$C, constructed in terms of the
three \SpR{3} irreps with probability  amplitudes defined by the
overlaps with the NCSM wavefunctions for $N_{max}=6$ case, were also used to determine
$B(E2~:~2^+_1~\rightarrow~0^+_{gs}~)$ transition rates. The latter, slightly increasing from $101\%$
to $107\%$ of the corresponding NCSM numbers with increasing \ho,
clearly reproduce the NCSM results. 

The focus here has been on demonstrating the existence of
\SpR{3} symmetry in NCSM results for $^{12}$C, and therefore a
possible path forward for extending the NCSM to a Sp-NCSM (symplectic no-core
shell model) scheme. This will allow one to account for even higher $\hbar\Omega$
configurations required to realize experimentally
measured B(E2) values without an effective charge, and especially
highly deformed spatial configurations required to reproduce
$\alpha$-cluster modes in heavier nuclei. In addition, the results can also be
interpreted as a further strong confirmation of Elliott's \SU{3} model since the projection of the
NCSM states onto the 0$\hbar\Omega$ space [Fig.
\ref{C12prblty_vs_hw}, blue (right) bars] is a projection of the NCSM results onto
the \SU{3} shell model. For example, for $^{12}$C the 0$\hbar\Omega$ \SU{3} symmetry 
ranges from just over 40\% of the NCSM $0^+_{gs}$ for $\hbar\Omega$ = 11 MeV to nearly 65\% for
$\hbar\Omega$ =18 MeV [Fig. \ref{C12prblty_vs_hw}, blue (left) bars] with 80\%-90\% of this symmetry
governed by the leading (0\,4) irrep.  
These numbers are consistent with what has been shown to be a dominance of the leading \SU{3}
symmetry for \SU{3}-based shell-model studies with realistic interactions in 0$\hbar\Omega$
model spaces.  It seems the simplest of Elliott's collective states can
be regarded as a good first-order approximation in the presence of
realistic interactions, whether the latter is restricted to a
0$\hbar\Omega$ model space or the richer multi-$\hbar\Omega$ NCSM model
spaces.

\section{Conclusions}
In summary, we demonstrated that {\it ab-initio} NCSM analysis starting 
with the JISP16 nucleon-nucleon interaction realize a collective
nucleon motion with a clear symplectic symmetry 
structure, which moreover remains unaltered whether the bare or effective interactions for
various \ho~ strengths are used. Specifically, NCSM wavefunctions for the lowest
$0^+_{gs}$, $2^+_1$ and $4^+_1$ states
in $^{12}$C and the ground state in $^{16}$O
project at the 85-90\% level onto very few \ph{0} and 2\ho~\ph{2}
spurious center-of-mass free symplectic irreps.
While the total dimensionality of the latter is only
$\approx 10^{-3}\%$ of the NCSM space, they also closely reproduce the NCSM $B(E2)$ estimates.
The results
confirm for the first time the validity of the \SpR{3} approach when
realistic interactions are invoked and when the most deformed 2\ho~\ph{2} 
symplectic irreps, which clearly improved the overlaps, are included. This demonstrates the
importance of the \SpR{3} symmetry, which simply matches the nuclear geometry to the many-nucleon
dynamics, as well as reaffirm the value of the simpler \SU{3} model upon which it is based. 

The results further suggest that a Sp-NCSM extension of the NCSM may be a
practical scheme for achieving convergence to measured B(E2) values
without the need for introducing an effective charge and even
for modeling cluster-like phenomena as these modes can be
accommodated within the general framework of the \SpR{3} model if
extended to large model spaces (high $N_{max}$). In addition, the symplectic extension of the
ab-initio NCSM that is ``structured" to take advantage
of massively parallel computing capability holds promise to allow us to model heavier nuclei
including neutron-deficient and $N\approx Z$ nuclei along the nucleosynthesis rp-path and unstable
nuclei currently explored in radioactive beam experiments. Heavy nuclei are also feasible due to the
natural extension of the \SpR{3} shell model to chiral-invariant pseudo-spin.

In short, the NCSM with a modern realistic $NN$
potential supports the development of collective motion in nuclei as
can be realized within the framework of the Sp-NCSM and as is
apparent in its 0$\hbar\Omega$ Elliott model limit. The Sp-NCSM is designed to model real nuclei
starting with realistic interactions (such as the ones based on
effective field theory), including, especially, momentum dependent forms.


\vskip 0.5cm
\centerline{\bf Acknowledgments}
The authors would like to thank Bruce Barrett and Andrey Shirokov for useful
discussions.
This work was supported by the US National Science Foundation, Grant
Nos 0140300 \& 0500291, and the Southeastern Universities Research
Association, as well as, in part, by the US Department
of Energy Grant Nos.  DE-AC02-76SF00515 and DE-FG02-87ER40371 and at the University of California,
Lawrence Livermore National Laboratory under contract No. W-7405-Eng-48. TD acknowledges supplemental
support from the Graduate School of Louisiana State University.

\end{document}